\documentclass[aps,prapplied,superscriptaddress,twocolumn]{revtex4-1}
\usepackage{bbm}
\usepackage{mathrsfs}
\usepackage{amsmath}
\usepackage{amsfonts}
\usepackage[colorlinks=true,linkcolor=blue,urlcolor=blue,citecolor=blue,anchorcolor=blue]{hyperref}
\usepackage{graphicx,epstopdf}
\usepackage{subfigure}
\usepackage{epsfig}
\usepackage{dcolumn}
\usepackage{bm}
\usepackage{color}
\usepackage{natbib}
\usepackage{amssymb}
\usepackage{xcolor}
\usepackage{braket}
\usepackage{ulem}
\usepackage{float}

\begin{document}
\title{Characterizing Multipartite Non-Gaussian Entanglement for Three-Mode Spontaneous Parametric Down-Conversion Process}

\author{Mingsheng Tian}
\affiliation{State Key Laboratory for Mesoscopic Physics, School of Physics, Frontiers Science Center for Nano-optoelectronics, Peking University, Beijing 100871, China}
\author{Yu Xiang}
\affiliation{State Key Laboratory for Mesoscopic Physics, School of Physics, Frontiers Science Center for Nano-optoelectronics, Peking University, Beijing 100871, China}
\affiliation{Collaborative Innovation Center of Extreme Optics, Shanxi University, Taiyuan, Shanxi 030006, China}
\author{Feng-Xiao Sun}
\email{sunfengxiao@pku.edu.cn}
\affiliation{State Key Laboratory for Mesoscopic Physics, School of Physics, Frontiers Science Center for Nano-optoelectronics, Peking University, Beijing 100871, China}
\affiliation{Collaborative Innovation Center of Extreme Optics, Shanxi University, Taiyuan, Shanxi 030006, China}
\author{Matteo Fadel}
\affiliation{Department of Physics, ETH Z\"urich, 8093 Z\"urich, Switzerland}
\affiliation{Department of Physics, University of Basel, Klingelbergstrasse 82, 4056 Basel, Switzerland}
\author{Qiongyi He}
\affiliation{State Key Laboratory for Mesoscopic Physics, School of Physics, Frontiers Science Center for Nano-optoelectronics, Peking University, Beijing 100871, China}
\affiliation{Collaborative Innovation Center of Extreme Optics, Shanxi University, Taiyuan, Shanxi 030006, China}
\affiliation{Peking University Yangtze Delta Institute of Optoelectronics, Nantong, Jiangsu, China}

\begin{abstract}
Very recently, strongly non-Gaussian states have been observed via a direct three-mode spontaneous parametric down-conversion in a superconducting cavity [Phys. Rev. X 10, 011011 (2020)].  The created multi-photon non-Gaussian correlations are attractive and useful for various quantum information tasks. However, how to detect and classify multipartite non-Gaussian entanglement has not yet been completely understood. Here, we present an experimentally practical method to characterize continuous-variable multipartite non-Gaussian entanglement, by introducing a class of nonlinear squeezing parameters involving accessible higher-order moments of phase-space quadratures. As these parameters can depend on arbitrary operators, we consider their analytical optimization over a set of practical measurements, in order to detect different classes of multipartite non-Gaussian entanglement ranging from fully separable to fully inseparable. We demonstrate that the nonlinear squeezing parameters act as an excellent approximation to the quantum Fisher information within accessible third-order moments. The level of the nonlinear squeezing quantifies the metrological advantage provided by those entangled states. Moreover, by analyzing the above mentioned experiment, we show that our method can be readily used to confirm fully inseparable tripartite non-Gaussian entangled states by performing a limited number of measurements without requiring full knowledge of the quantum state.
\end{abstract}

\maketitle

\section{Introduction} 

Continuous-variable (CV) systems, where multimode entangled states can be deterministically prepared~\cite{weedbrook2012gaussian,yokoyama2013ultra,roslund2014wavelength,chen2014experimental,armstrong2015multipartite,cavalcanti2015detection,cai2017multimode,deng2017demonstration,larsen2019deterministic,asavanant2019generation,takeda2019demand,cai2020versatile,wang2020deterministic}, constitute an important platform for quantum technologies, including quantum teleportation networks~\cite{PhysRevLett.80.869,nature.2004}, quantum key distribution~\cite{PhysRevLett.88.057902}, quantum secrete sharing~\cite{armstrong2015multipartite}, boson sampling~\cite{PhysRevLett.119.170501,np.15.925}, and multi-parameter quantum metrology~\cite{np.16.3,PhysRevLett.121.130503}. Non-Gaussian states in CV systems have attracted increasing attentions in recent years~\cite{np11.713,PRXQuantum.2.030204}, as they have been proven to be indispensable resources for universal quantum computation~\cite{niset2009no,PhysRevLett.109.230503}, entanglement distillation~\cite{eisert2002distilling,fiuravsek2002gaussian,nphoton.2010.1}, quantum-enhanced sensing~\cite{science.345,PhysRevLett.107.083601}, and quantum imaging~\cite{PhysRevApplied.16.064037}. Such perspectives have led to a growing interest in the experimental preparation of multimode non-Gaussian quantum states~\cite{natphy.nicolas,olthreemode,prxthreemode}. 

Despite the fact that substantial progress has been made in the generation of multimode non-Gaussian states, the characterization of their entanglement structure still poses a number of conceptual and practical challenges. The main reason being that the nontrivial correlations appear in higher-order moments of the observables that cannot be sufficiently uncovered by the widely used entanglement criteria based on second-order correlations~\cite{PhysRevLett.84.2726,PhysRevA.67.052315,PhysRevA.72.032334,PhysRevLett.96.050503,PhysRevLett.111.250403,PRA.90.062337}, partially hindering their application for quantum information tasks. To tackle this problem, several approaches have been developed to take higher-order moments into account, such as the entropic entanglement criteria~\cite{PhysRevLett.103.160505,RevModPhys.90.035007,PhysRevA.103.013704}, the generalized Hillery-Zubairy criteria based on the multimode moments~\cite{PhysRevA.81.062322,PhysRevA.84.032115,PhysRevLett.125.020502}, or nonlinear entanglement criteria based on the amplitude-squared squeezing~\cite{PhysRevLett.114.100403,PhysRevLett.127.150502}. The quantum Fisher information (QFI) also provides a powerful method for capturing strongly non-Gaussian features of quantum states. It is widely applied to detect multi-particle entanglement in nonclassical spin states~\cite{rev.mod.phys.90.035005, science.345,npys3700,PhysRevA.102.012412,PhysRevLett.126.080502,NC2021,fadel2022multiparameter} and has been recently developed to characterize continuous variables~\cite{npj.5.2056,pra2016manuel,Gessner2017entanglement}. Moreover, QFI provides a powerful tool to establish a quantitative link between entanglement and quantum metrology~\cite{PhysRevLett.102.100401,pra.85.022321}.
However, most of the above mentioned methods are experimentally challenging, as they require full knowledge of the quantum state, which is extremely difficult in the multipartite scenario. Recently, a nonlinear squeezing parameter was proposed for the metrological characterization of non-Gaussian features~\cite{PhysRevLett.122.090503,guo2021detecting}. This can be optimized over a set of observables that are experimentally accessible, and it can ultimately coincide with the state QFI with increasing the order of the measured moments. 

Experimentally, the recent advance in the generation of three-photon correlated non-Gaussian states via a direct three-mode spontaneous parametric down-conversion (SPDC) process~\cite{prxthreemode}, opens up novel possibilities for various quantum information processing applications. Thus, it would be interesting to develop experimentally feasible methods to witness different classes of multipartite entanglement in such systems. 

In this paper, we construct a class of CV nonlinear squeezing parameters for testing fully separable, inseparable, and fully inseparable tripartite non-Gaussian states with practical measurements. Firstly, we analyze the QFI for arbitrary sets of accessible local observables (involving higher-order moments) in three subsystems and find out the optimal combination that maximizes the violation of different entanglement bounds. These bounds are divided into three classes according to the separability properties with respect to the three splittings (fully separable) and particular bipartite splittings (biseparable). To avoid the requirement of full quantum state tomography, we then construct a nonlinear squeezing parameter by analytically determining the optimal measurements within arbitrary accessible third-order observables, that results in an excellent approximation to the QFI. The level of the nonlinear squeezing parameter quantifies the metrological advantage provided by different classes of entangled states. Moreover, by analyzing the accessible conditions in the three-mode SPDC experiment of Ref.~\cite{prxthreemode}, we show that our method is capable of detecting fully inseparable tripartite entangled states by performing a limited number of measurements. Our results lead to an experimentally feasible way to systematically investigate CV multipartite non-Gaussian entanglement, which paves a way for exploiting their potential quantum advantages beyond Gaussian states. 

\section{A general method to detect non-gaussian entanglement}
Before considering a specific system, we firstly introduce a general method to detect non-Gaussian entanglement with a two-step optimization.

\textit{Step 1.$-$} 
We characterize the entanglement with choosing optimal operators in QFI. 
For an arbitrary $N$-partite separable quantum state $\hat{\rho}_{\mathrm{sep}}$, it was shown that the QFI must be less than a bound $B_n$ given by the variance~\cite{pra2016manuel}
\begin{equation}\label{eqfv}
F_Q\left[\hat{\rho}_{\mathrm{sep}},  \sum_{j=1}^N \hat{A}_j\right] \leq 4 \sum_{j=1}^{N} \mathrm{Var}\left(\hat{A}_{j}\right)_{\hat{\rho}_{\mathrm{sep}}} \equiv B_n.
\end{equation}
Here, $\hat{A}_j$ is a local observable in the reduced state $\hat{\rho}_j$, and $\mathrm{Var}(\hat{A}_j)_{\hat{\rho}}=\langle \hat{A}_j^2\rangle_{\hat{\rho}}-\langle \hat{A}_j\rangle^2_{\hat{\rho}}$ indicates the variance. $F_Q$ denotes the QFI, which describes the sensitivity of the parameter $\theta$ when the state $\hat{\rho}$ is transformed with unitary evolution $\hat{\rho}_{\theta}=e^{-i \sum_j \hat{A}_j \theta} \hat{\rho} e^{i \sum_j \hat{A}_j \theta}$ and provides a bound on the accuracy to determine $\theta$ as $(\Delta\theta_{est})^2\geq1/F_Q[\hat{\rho},\sum_j \hat{A}_j ]$.

Since Eq.~(\ref{eqfv}) represents a necessary criterion for separability, its violation is sufficient criterion for entanglement.
In order to witness the entanglement in the largest possible parameter range, we need to choose an optimal local operator $\hat{A_j}$, 
which can be constructed by analytical optimizing over arbitrary linear combination of accessible operators (namely, $\hat{A}_j=\sum_{m=1}c_j^{(m)}\hat{A}_j^{(m)}=\mathbf{c}_j\cdot\mathbf{\hat{A}}_j$).
In this case, the full operator
$\hat{A}(\mathbf{c})=\sum_{j=1}^{N} \mathbf{c}_{j} \cdot \hat{\mathbf{A}}_{j}$
is characterized by the combined vector $\mathbf{c}=(\mathbf{c}_1,\cdots,\mathbf{c}_N)^T$.
According to Eq.~(\ref{eqfv}), the quantity
$
W[\hat{\rho},\hat{A}(\mathbf{c})]=F_Q[\hat{\rho},\hat{A}(\mathbf{c})]-4\sum_{j=1}^N \mathrm{Var}(\mathbf{c}_j\cdot\mathbf{\hat{A}}_j)_{\hat{\rho}}
$
must be nonpositive for arbitrary choices of $\mathbf{c}$ whenever the state is separable. 
We can now maximize $W[\hat{\rho},\hat{A}(\mathbf{c})]$ by variation of $\mathbf{c}$ to obtain an optimized entanglement operator $\hat{A}$. 
According to Ref.~\cite{pra2016manuel}, the quantity can be expressed as $W[\hat{\rho},\hat{A}(\mathbf{c})]=\mathbf{c}^{T}\left(Q_{\hat{\rho}}^{\mathcal{A}}-4 \Gamma_{\Pi(\hat{\rho})}^{\mathcal{A}}\right) \mathbf{c}$ and the optimal $\mathbf{c}$ can be obtained by calculating the eigenvector of the maximum eigenvalue of the matrix $Q_{\hat{\rho}}^{\mathcal{A}}-4 \Gamma_{\Pi(\hat{\rho})}^{\mathcal{A}}$
(see Appendix \ref{qfi} for details).

\textit{Step 2.$-$} 
Computing the QFI is challenging for arbitrary multipartite states, as it requires the full density matrix of the system~\cite{PhysRevLett.72.3439}. For this reason, it is often more practical to find its lower bound, which, regarded as squeezing parameters $\chi^2$, involves simple measurements that are usually experimentally feasible. The relation between the QFI and $\chi^2$ fulfills~\cite{PhysRevLett.102.100401},
\begin{equation}\label{eqfangsuo}
F_{\mathrm{Q}}[\hat{\rho}, \hat{A}] 
\geq \frac{\left|\langle[\hat{A}, \hat{M}]\rangle_{\hat{\rho}}\right|^{2}}{\mathrm{Var}(\hat{M})_{\hat{\rho}}}
\equiv \chi^{-2}(\hat{\rho},\hat{A},\hat{M}),
\end{equation}
which is saturable for certain measurement operator $\hat{M}$. 
Often $\hat{M}$ is a complicated operator, of difficult implementation. Therefore, similarly to the analytical optimization of $\hat{A}$, $\hat{M}$ will be optimized over a linear combination of experimentally practical observables (see the details in Appendix~\ref{optsqueezing}). 

It is noticed that the general method introduced here is able to be applied to arbitrary quantum states for detecting the multipartite entanglement, where the optimal operator $\hat{A}$ and $\hat{M}$ will differ for different systems. 
In the following, we will take a particular non-Gaussian system as an example and obtain the optimal operator $\hat{A}$ and $\hat{M}$ for detecting the entanglement with the above two-step optimization.

\section{Witnessing entanglement in three-mode non-Gaussian states via the QFI} 

Here, we focus on the three-mode SPDC process realized in Ref.~\cite{prxthreemode}, where a pump photon at frequency $\omega_p$ is down converted to three nondegenerate photons at frequencies $\omega_1,~\omega_2,~\omega_3$, respectively. This process is described by the interaction Hamiltonian
\begin{equation}\label{eqham}
    \hat{H}=i\hbar \kappa ({\hat{b}}\hat{a}_1^{\dagger}\hat{a}_2^{\dagger}\hat{a}_3^{\dagger}-{\hat{b}}^{\dagger}\hat{a}_1\hat{a}_2\hat{a}_3),
\end{equation}
where $\kappa$ is the third-order coupling constant, $\hat{b}$ and $\hat{a}_i$ ($i=1,2,3$) are the annihilation operators for the pump and $i$th mode, respectively. 
With this Hamiltonian, the three-mode non-Gaussian state is obtained by {$\partial \hat{\rho}/\partial t=-i[\hat{H},\hat{\rho}]/\hbar$} with considering the initial state as vacuum for the generated triplets and coherent mode $\alpha_p$ for the pump.

A three-mode fully separable state allows for a description in three splittings $1|2|3$ ($\hat{\rho}_{\mathrm{sep}}=\sum_{k} p_{k} \hat{\rho}_{1}^{k} \otimes \hat{\rho}_{2}^{k} \otimes \hat{\rho}_{3}^{k}$), which provides the bound $B_1$. Thus, observing $F_Q > B_1$ indicates inseparability among the three modes. Furthermore, we can denote the maximum among the sum of variances for different bipartitions $1|23, ~2|13, ~3|12$, as bound $B_2$. The state is then confirmed to be fully inseparable if $F_Q > B_2$ (see Appendix \ref{optfisher} for details).  
\begin{figure}[tbp]
	\centering
	\includegraphics[scale=0.42]{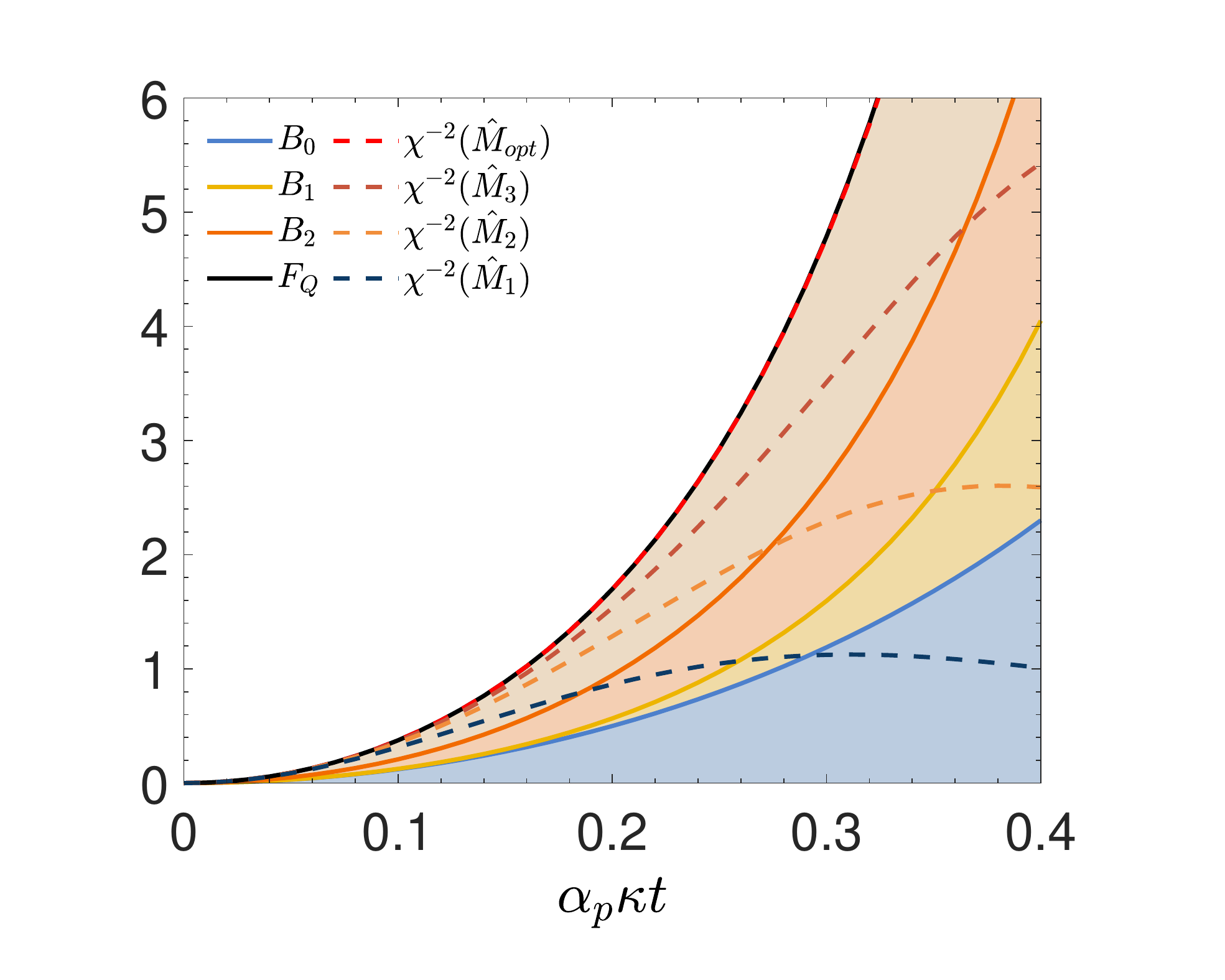}
	\caption{The QFI $F_Q[\hat{\rho},\hat{N}]$, the nonlinear parameters $\chi^{-2}$, amd the bounds for different classes of three-mode quantum states as functions of the effective coupling strength $\alpha_p \kappa t$ created by a direct three-mode SPDC process. The lowest solid curve represents the metrological bound $B_0$, the highest solid curve is the QFI $F_Q[\hat{\rho},\hat{N}]$, and the two solid curves between them (from bottom to top) are the fully separable bound $B_1$, and the biseparable bound $B_2$, respectively. The different dashed lines are obtained by considering different operators $\hat{M}_i$ in the nonlinear squeezing parameters $\chi^{-2}(\hat{M}_i)$.
	}
	\label{fig1}
\end{figure}

A key step to witness tripartite non-Gaussian entanglement is the choice of the local observables $\hat{A}_j$, corresponding to the generators of parameter imprinting. To find the optimal operator $\hat{A}_j$, we express $\mathcal{A}=\{\hat{\mathbf{A}}_{1},\hat{\mathbf{A}}_{2}, \hat{\mathbf{A}}_{3}\}$ through a set of accessible local operators for each mode $\hat{\mathbf{A}}_{j}=\{\hat{A}_j^{(1)},\hat{A}_j^{(2)},\dots\}$. The optimization of local observable $\hat{A}_{opt}$ to witness entanglement is analyzed over arbitrary linear combinations of the accessible operators. As the generated states are non-Gaussian, their characteristics cannot be sufficiently captured by linear quadratures $\hat{x}_j,\hat{p}_j$, where  $\hat{x}_j=\hat{a}_j+\hat{a}_j^{\dagger}$,  $\hat{p}_j=-i(\hat{a}_j-\hat{a}_j^{\dagger})$. Therefore, we extend the family of accessible operators to the second-order, i.e. $\mathbf{\hat{A}}_j=[\hat{x}_j,\hat{p}_j,\hat{x}_j^2,\hat{p}_j^2,(\hat{x}_j\hat{p}_j+\hat{p}_j\hat{x}_j)/2]$. 

To find the optimal linear combination, we make use of the matrix form $W[\hat{\rho}, \hat{A}(\mathbf{c})]=\mathbf{c}^{T}\left(Q_{\hat{\rho}}^{\mathcal{A}}-4 \Gamma_{\Pi(\hat{\rho})}^{\mathcal{A}}\right) \mathbf{c}$ with the non-Gaussian state $\hat{\rho}$ generated in the three-mode SPDC process. Then by calculating the maximum eigenvalue and eigenvectors of the matrix $Q_{\hat{\rho}}^{\mathcal{A}}-4 \Gamma_{\Pi(\hat{\rho})}^{\mathcal{A}}$, the optimal local operator $\hat{A}_{opt}=\sum_{j=1}^3(x_j^2+p_j^2)$ is obtained. It can also be rewritten as $\hat{A}_{opt}=\sum_{j=1}^3\hat{a}_j^{\dagger}\hat{a}_j=\hat{N}$, because the constant term in $\hat{A}_{opt}$ has no effects on the results (see the detailed analytical optimization in Appendix~\ref{optfisher}).

By comparing the QFI $F_Q[\hat{\rho},\hat{A}_{opt}]$ to different bounds, such as the fully separable bound $B_1$ and the biseparable bound $B_2$, the three-mode non-Gaussian entanglement can be classified. In Fig.~\ref{fig1}, we plot the QFI and the different bounds (solid curves) as functions of the effective coupling strength $\alpha_{p}\kappa t$, where $t$ denotes the effective interaction time. We find that the QFI $F_Q[\hat{\rho},\hat{A}_{opt}]$ is higher than the biseparable bound $B_2$, which means that the non-Gaussian states generated in the three-mode SPDC dynamics Eq.~(\ref{eqham}) are fully inseparable.

It has been clarified that only a specific class of entangled states enable quantum-enhanced parameter estimation~\cite{rev.mod.phys.90.035005}, where a sufficient condition for quantum-enhanced metrology is~\cite{RN612},
 \begin{equation}\label{eqclassical}
   F_Q [\hat{\rho},\hat{N}] > 4 \text{Tr}[\hat{\rho}\hat{N}] \equiv B_0,
 \end{equation}
where $\hat{N}=\sum_j^3\hat{a}_j^{\dagger}a_j$ is the total number operator for the three modes. The right hand side of the inequality give a bound $B_0$ for QFI (the derivation is given in Appendix~\ref{C}). It is obviously seen from Fig.~\ref{fig1} that both of the entanglement bounds ($B_1$ and $B_2$) are above the metrological bound $B_0$ (the lowest solid curve), which indicates that the entanglement witnessed in the three-mode SPDC dynamics is useful for quantum metrology.

\section{Detecting non-Gaussian entanglement with nonlinear squeezing parameters}

In this section, we will take the \textit{Step 2} optimization to obtain the squeezing parameter $\chi^2$ with optimal measurement $\hat{M}$. By extending the family of accessible observables to third order, the optimal $\hat{M}$ we obtained is $\hat{M}_{opt}=\hat{x}_1 \hat{x}_2 \hat{x}_3-\hat{x}_1 \hat{p}_2 \hat{p}_3-\hat{p}_1 \hat{x}_2 \hat{p}_3-\hat{p}_1 \hat{p}_2 \hat{x}_3$ (see the details in Appendix~\ref{optsqueezing}). As shown in Fig.~\ref{fig1} by the red dashed curve, the nonlinear squeezing parameter with optimal choices of measurements $\chi^{-2}(\hat{\rho},\hat{A}_{opt},\hat{M}_{opt})$ approximates extremely well the exact QFI $F_{\mathrm{Q}}[\hat{\rho}, \hat{A}_{opt}]$. Hence, the squeezing parameter $\chi^{-2}(\hat{\rho},\hat{A}_{opt},\hat{M}_{opt})$ can act as a faithful substitute for the QFI in the left-hand side of inequalities (\ref{eqfv}) and (\ref{eqclassical}), avoiding to perform quantum state tomography.
\begin{figure}[tbp]
    \centering
    \includegraphics[scale=0.42]{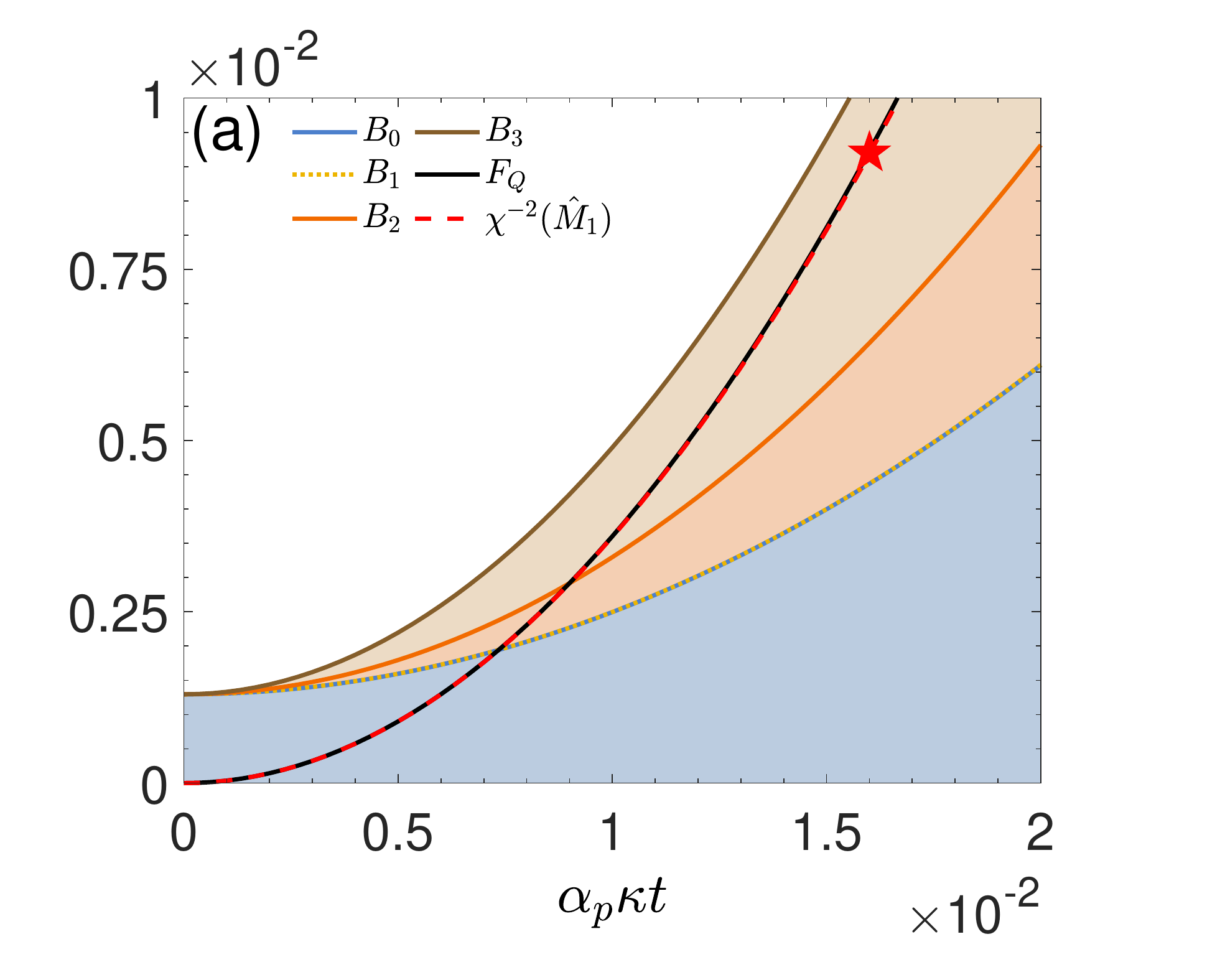}\\
    \includegraphics[scale=0.42]{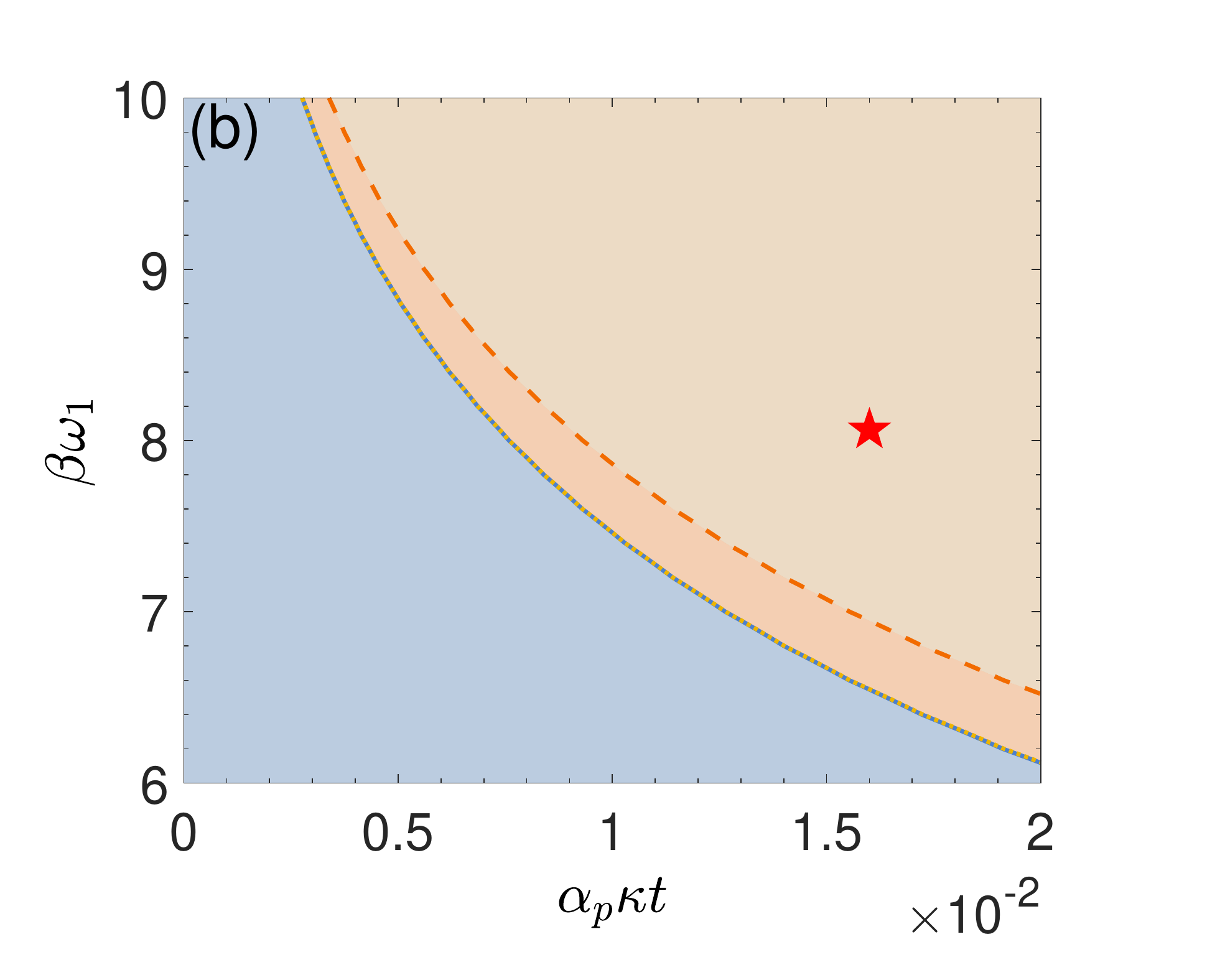}
    \caption{(a) The QFI $F_Q$, the nonlinear parameter $\chi^{-2}(\hat{M}_1)$, and the bounds of different separability classes versus $\alpha_p\kappa t$, for $T=25$ mK. The bound $B_3$ is satisfied by all quantum states (see Appendix~\ref{optfisher}). (b) The condition to witness fully inseparable, biseparable entanglement, and metrological advantage consists in observing a nonlinear parameter $\chi^{-2}(\hat{M}_1)$ that is higher than $B_2$ (dashed orangered), $B_1$ (dotted yellow) and $B_0$ (solid blue), respectively. These bounds vary with the effective coupling strength $\alpha_{p}\kappa t$ and inverse temperature $\beta \omega_1=\hbar \omega_1/(k_BT)$. The red star denotes the experimentally generated state with $T=25$ mK and $\alpha_p\kappa t \approx 0.016$ reported in \cite{prxthreemode}, located in the area of fully inseparable tripartite entangled states.
}
    \label{fig2}
\end{figure}

Although more accessible in experiments comparing to the QFI, the nonlinear squeezing parameter $\chi^{2}(\hat{\rho},\hat{A}_{opt},\hat{M}_{opt})$ can still be of demanding implementation, as (for the three-mode SPDC process) it requires measuring four collective observables in $\hat{M}_{opt}$. Therefore, we show that this can be further simplified by a measurement $\hat{M}$ which is still capable of detecting fully tripartite inseparable states but contains less number of observables. For this purpose, an experimentally feasible coincidence measurement is evaluated, which can be written as $\hat{Q}=[\hat{x}_1\sin(\theta_1)+\hat{p}_1\cos(\theta_1)]\times[\hat{x}_2\sin(\theta_2)+\hat{p}_2\cos(\theta_2)]\times[\hat{x}_3\sin(\theta_3)+\hat{p}_3\cos(\theta_3)]$, where $\theta_j\in [0,2\pi)$. The alternative observable $\hat{M}$ can be obtained by varying $\theta_j$. For the simplest case of $\hat{M_1}$ that contains only one collective observable, by optimizing $\theta_{1,2,3}$ we find that the nonlinear parameter $\chi^{-2}(\hat{M}_1)=|\langle[\hat{A}_{opt}, \hat{M}_1]\rangle_{\hat{\rho}}|^{2}/\mathrm{Var}( \hat{M}_1)_{\hat{\rho}}$ is maximized by any one of four terms in $\hat{M}_{opt}$, e.g. $\hat{M}_1=\hat{x}_1 \hat{x}_2 \hat{x}_3$. 
We find that this performs as an excellent proxy for the QFI when $\alpha_{p}\kappa t\leq 0.1$, as indicated by the blue dashed line in Fig.~\ref{fig1}. 

To test entanglement, we need to compare the simplified nonlinear parameter $\chi^{-2}(\hat{M}_1)$ with different bound $B_n$. The simplified nonlinear parameter has the form of,
\begin{equation}
\begin{aligned}
\chi^{-2}(\hat{M}_1)
&=\frac{|\langle[\hat{A}_{opt}, \hat{M}_1]\rangle_{\hat{\rho}}|^{2}}
{\mathrm{Var}(\hat{M}_1)_{\hat{\rho}}}\\
&=\frac{|\langle
\hat{p}_1\hat{x}_2\hat{x}_3+\hat{x}_1\hat{p}_2\hat{x}_3+\hat{x}_1\hat{x}_2\hat{p}_3
\rangle_{\hat{\rho}}|^{2}}
{\mathrm{Var}(\hat{x}_1\hat{x}_2\hat{x}_3)_{\hat{\rho}}}.
\end{aligned}
\end{equation}
In experiments, $\chi^{-2}(\hat{M}_1)$ can be obtained by detecting $\hat{p}_1\hat{x}_2\hat{x}_3$, $\hat{x}_1\hat{p}_2\hat{x}_3$, $\hat{x}_1\hat{x}_2\hat{p}_3$, and  $\hat{x}_1\hat{x}_2\hat{x}_3$ with collective measurements~\cite{prxthreemode,pan1999,sa.collective.measurement,npj.collective.measurement}.
And the different bounds $B_n=4 \sum_{j} \mathrm{Var}(\hat{N}_{j})$ can be obtained by  number-resolving detectors~\cite{np2007-Photon-number-discriminating,np2008-detector,np-superconducting}.

Besides, we also provide the results of simplified $\hat{M}_{2}$ and $\hat{M}_{3}$ that involve two and three collective observables, receptively, as shown by the two middle dashed curves in Fig.~\ref{fig1} (the detailed expressions and analysis can be found in Appendix~\ref{simplesqueezing}). As expected, the more observables are included in $M$, the larger range of coupling strengths over which non-Gaussian multipartite entanglement can be witnessed.

\section{Testing fully inseparable tripartite entangled states in three-photon SPDC experiment} 

In this section we use the QFI $F_Q$ and the nonlinear parameter $\chi^{-2}$ to examine the three-mode non-Gaussian entanglement for a realistic scenario. In experiments, thermal noise in the initial state $\hat{\rho}_j(n_{th}^j)$ is inevitable, with average thermal photon number $\langle n_{th}^j \rangle=1/(e^{\beta\omega_j}-1)$, and  $\beta \omega_j=\hbar\omega_j/(k_B T)$. Here, we refer to the reported tripartite states produced by three-mode SPDC in the superconducting experiment~\cite{prxthreemode}, where the noise temperature is $T=25$ mK and the frequencies of three modes are $\omega_1=2\pi\times 4.2$GHz, $\omega_2=2\pi\times 6.1$GHz, $\omega_3=2\pi\times 7.5$GHz, respectively. In this case, the simplified squeezing parameter $\chi^2(\hat{\rho},\hat{A}_{opt},\hat{M}_1)$ is sufficient to characterize the entanglement, since the achieved effective coupling strength is quite small. 

Specifically, the experimental result for $\alpha_p \kappa t \approx 0.016$ is marked as a red star in Fig.~\ref{fig2}, where the parameters are verified by reproducing the reported results in Ref.~\cite{prxthreemode}. As shown in Fig.~\ref{fig2}(a), the simplified nonlinear parameter $\chi^{-2}(\hat{\rho},\hat{A}_{opt},\hat{M}_1)$ (red dashed line) approximates well the QFI $F_Q(\hat{\rho},\hat{A}_{opt})$ (solid black line). Thus, $\chi^{-2}(\hat{\rho},\hat{A}_{opt},\hat{M}_1)$ is adequate to witness entanglement in such range of coupling strengths. Besides this, different separable bounds to characterise entanglement are illustrated in the figure, showing that the metrological bound $B_0$ coincides with the separable bound $B_1$ in this small coupling regime.
\begin{figure}[t]
	\centering
	\includegraphics[scale=0.42]{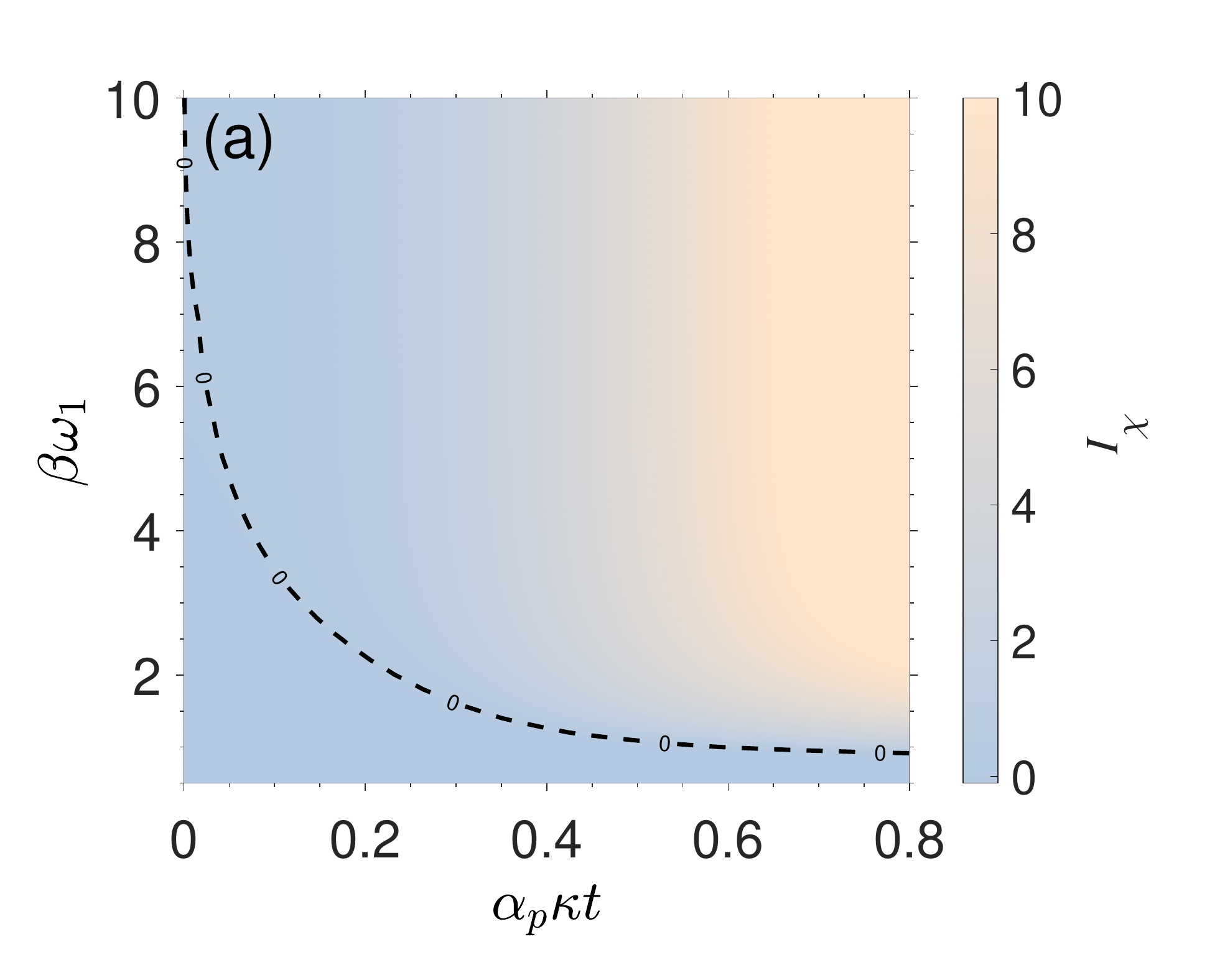}\\
	\includegraphics[scale=0.42]{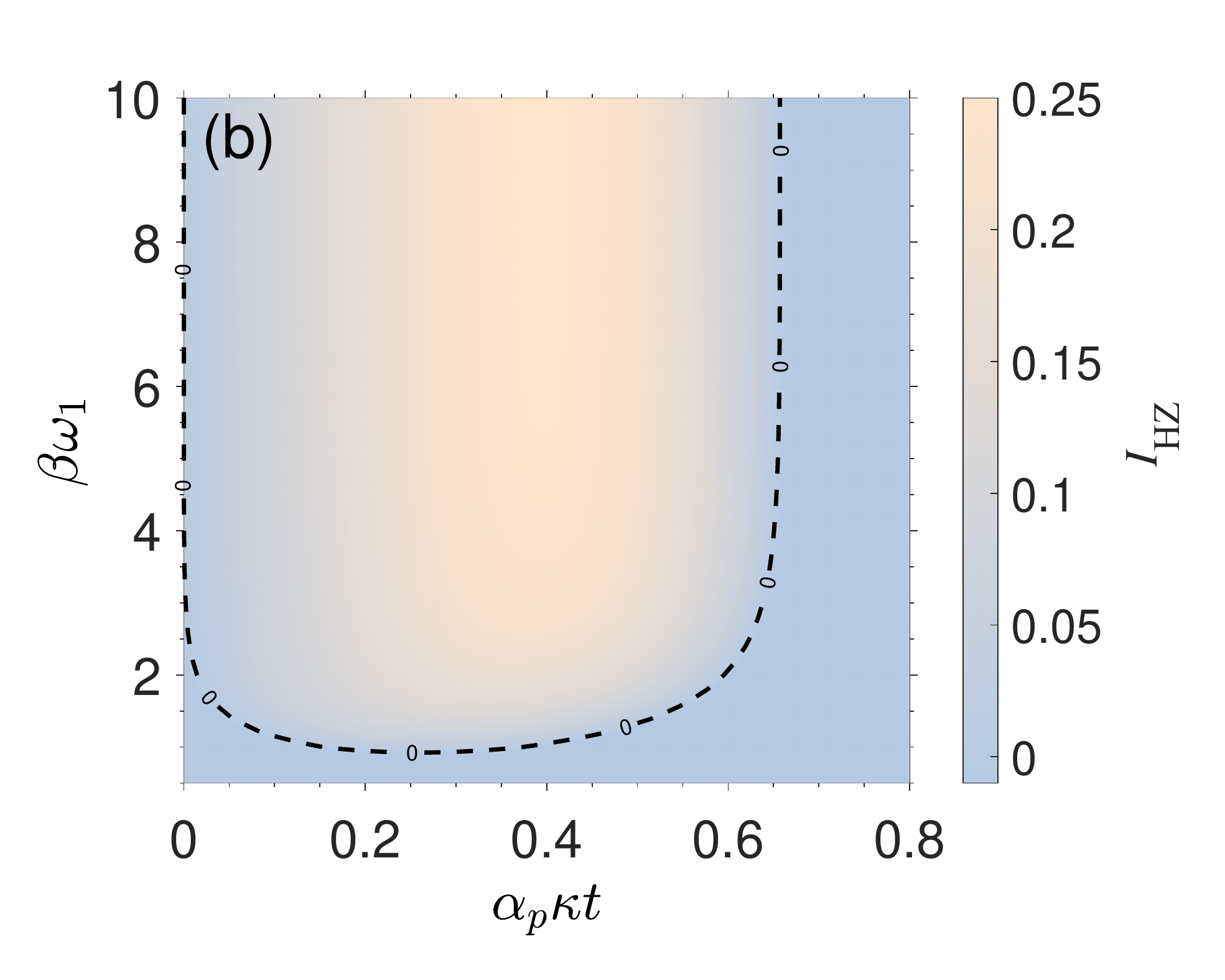}
    \caption{The values of witness (a) $I_{\chi}$ and (b) $I_\mathrm{HZ}$ as functions of both effective coupling strength $\alpha_{p}\kappa t$ and inverse temperature $\beta\omega_1$, when system evolves under Hamiltonian (\ref{eqham}) for three modes with $\omega_1=2\pi\times 4.2$GHz, $\omega_2=2\pi\times 6.1$GHz, $\omega_3=2\pi\times 7.5$GHz. A positive value of $I_\chi$ or $I_\mathrm{HZ}$ (the region above the dashed curve) indicates full inseparability of the three modes.}
	\label{fig3}
\end{figure}

In summary, Fig.~\ref{fig2}(b) shows that we can characterize the entanglement of the final state over a wide range of both coupling strength and environmental temperature. Full tripartite inseparability is witnessed with larger coupling and lower temperature (top right area) by violating the bound $B_2$ of all three possible $i|jk$ bipartitions. The intermediate area, where $\chi^2(\hat{\rho},\hat{A}_{opt},\hat{M}_1)$ violates the tripartite $a|b|c$ bound $B_1$ but doesn't violate $B_2$, indicates there is entanglement among three modes. States in the bottom left area are not detected as useful for quantum metrology, since $\chi^2(\hat{\rho},\hat{A}_{opt},\hat{M}_1)$ is lower than the metrological bound $B_0$. In addition, as the red star is located in the top right area, a metrological useful full tripartite inseparability is witnessed in the experimentally reported three-mode states~\cite{prxthreemode}.

\section{Comparison to other criteria}

In this section we compare our criterion to the widely used Hillery-Zubairy criterion~\cite{PhysRevA.81.062322}, where full inseparability is witnessed if $I_\mathrm{HZ}=\mathrm{min}\{I_1,I_2,I_3\}>0$, where $I_i=|\langle\hat{a}_1\hat{a}_2\hat{a}_3\rangle|-\sqrt{\langle \hat{N}_i\rangle \langle \hat{N}_j \hat{N}_k\rangle}$ and $\hat{N}_i=\hat{a}_i^{\dagger}\hat{a}$, $i\neq j\neq k\neq i$.
Similarly, we define $I_{\chi}=\chi^{-2}(\hat{\rho},\hat{A}_{opt},\hat{M}_{opt})-B_2$ based on the optimal measurements we proposed, such that a state is fully inseparable if $I_{\chi}>0$. The values of $I_\mathrm{HZ}$ and $I_{\chi}$ are presented in Fig.~\ref{fig3} as functions of the effective coupling strength and the inverse temperature, where the three-mode states are generated under the interaction Hamiltonian (\ref{eqham}). It is shown that $I_\mathrm{HZ}$ fails to detect tripartite inseparability when $\alpha_{p}\kappa t>0.66$, while our witness $I_{\chi}$ works well in a larger coupling regime. This can be explained by the fact that detecting entanglement for large coupling strength requires observing correlations of higher order, which are not considered in the witness $I_\mathrm{HZ}$. 

As it is shown in Fig.~\ref{fig3}(a), $I_{\chi}$ fails to witness fully inseparable tripartite entangled states in the area below the dashed black line at the region with very weak coupling and high temperature, where $I_\mathrm{HZ}$ has a better performance [Fig.~\ref{fig3}(b)]. This result can be explained by the fact that our entanglement conditions are tailored to detect metrologically useful entanglement, and that the region where the HZ criterion outperforms our criterion might correspond to states less useful for parameter-estimation tasks.

\section{Conclusions} 

Based on the widely used quantifiers for metrological sensitivity, we have obtained an experimentally practical entanglement witness for CV tripartite non-Gaussian states generated by three-mode SPDC. First, the optimal phase-space observable that maximizes the QFI has been analyzed to detect the tripartite non-Gaussian entanglement, where fully inseparable states, biseparable entangled states, and fully separable states can be distinguished. Then, in order to avoid the full quantum state tomography required by the QFI, we constructed a nonlinear squeezing parameter by analytically determining the optimal measurements within arbitrary accessible observables, and demonstrate that it performs as well as the QFI by involving up to third-order moments of phase-space measurements. Moreover, by considering the experimental conditions in a recent three-mode SPDC experiment, we show that our method can detect fully inseparable tripartite non-Gaussian entanglement with only a limited number of measurements. Notably, the level of the nonlinear squeezing parameter quantifies the metrological advantage provided by the examined entangled states. Our results provide an approach to understand and characterize multipartite non-Gaussian entanglement, and paves the way to harness their potential applications in quantum metrology experiments.

\begin{acknowledgments}

This work is supported by the National Natural Science Foundation of China (Grants No.~11975026, No.~12125402, and No.~12147148), Beijing Natural Science Foundation (Grant No.~Z190005), the Key R\&D Program of Guangdong Province (Grant No. 2018B030329001), and the LabEx ENS-ICFP: ANR-10-LABX-0010 / ANR-10-IDEX-0001-02 PSL*. F.-X. S. acknowledges the China Postdoctoral Science Foundation (Grant No. 2020M680186).
\end{acknowledgments}


\appendix

\section{Quantum Fisher Information}\label{qfi}

\subsection{Fisher Information} \label{fisher}

Originally, the Fisher information was introduced in the context of parameter estimation (see Ref.~\cite{RN01} for a review). To infer the value of $\theta$, one performs a measurement $\hat{M}=$ $\left\{\hat{M}_{\mu}\right\}$, which in the most general case is given by a positive operator valued measure (POVM). The Fisher information $F[\hat{\rho}(\theta), \hat{M}]$ quantifies the sensitivity of n independent measurements and gives a bound on the accuracy to determine $\theta$ as $(\Delta \theta)^{2} \geq 1 / (nF[\hat{\rho}(\theta), \hat{M}])$ in central limit ($n\gg1$). In particular, the Fisher information is defined as \cite{PhysRevLett.72.3439}
\begin{equation}
    F[\hat{\rho}(\theta), \hat{M}] \equiv \sum_{\mu} \frac{1}{P(\mu \mid \theta)}\left(\frac{\partial P(\mu \mid \theta)}{\partial \theta}\right)^{2},
\end{equation}
where $P(\mu \mid \theta) \equiv \operatorname{Tr}\left\{\hat{\rho}(\theta) M_{\mu}\right\}$ is the probability to obtain the measurement outcome $\mu$ in a measurement of $\hat{M}$ given the state $\rho(\theta).$

The Fisher information for an optimal measurement, i.e., the one that gives the best resolution to determine $\theta$, is called quantum Fisher information (QFI), and is defined as $F_{Q}[\hat{\rho}(\theta)] \equiv$ $\max _{\hat{M}} F[\hat{\rho}(\theta), \hat{M}] .$ There, one is interested in distinguishing the state $\hat{\rho}$ from the state $\hat{\rho}({\theta})=e^{-i \hat{A} \theta} \hat{\rho} e^{i \hat{A} \theta}$, obtained by applying an unitary induced by a Hermitian generator $\hat{A}$.  With the spectral decomposition $\hat{\rho}=\sum_k p_k|\Psi_k\rangle \langle \Psi_k|$, an explicit expression for $F_Q[\hat{\rho},\hat{A}]$ is given by \cite{PhysRevLett.72.3439}
\begin{equation}
F_{Q}[\hat{\rho}, \hat{A}]=2 \sum_{\substack{k, l\\p_k+p_l\neq 0}} \frac{\left(p_{k}-p_{l}\right)^{2}}{p_{k}+p_{l}}\left|\left\langle\Psi_{k}\right|\hat{A}\left| \Psi_{l}\right\rangle\right|^{2}.
\end{equation}
And in pure states, it takes the simple form $ F_{Q}[|\psi\rangle\langle\psi|, \hat{A}]=4(\Delta A)^{2}.  $

\subsection{The Optimal operator in QFI}\label{optfisher}

Any seperable quantum state 
$\hat{\rho}_{\mathrm{sep}}=\sum_{n} p_{n} \hat{\rho}_{1}^{n} \otimes \cdots \otimes \hat{\rho}_{N}^{n}$ must satisfy~\cite{pra2016manuel}
\begin{equation}\label{eqfv1}
F_Q\left[\hat{\rho}_{\mathrm{sep}},  \sum_{j=1}^N \hat{A}_j\right] \leq 4 \sum_{j=1}^{N} \mathrm{Var}\left(\hat{A}_{j}\right)_{\hat{\rho}_{\mathrm{sep}}}=B_n,
\end{equation}
where $\hat{A}_i$ is the local operator acting on the reduced state $\hat{\rho}_i$. $\mathrm{Var}(A)_{\hat{\rho}}=\langle A^2\rangle_{\hat{\rho}}-\langle A\rangle^2_{\hat{\rho}}$ denotes the variance. If a state violates this inequality, it can not be divided into the corresponding seperable quantum state. In our case, for a generated state with the local operators $\hat{A}$, the right-hand side of Eq.~(\ref{eqfv1}) can be characterized by the different bounds, taking the form of
\begin{align}
    B_1=4\, [\mathrm{Var}(&\hat{A}_1)_{\hat{\rho}_a}+\mathrm{Var}(\hat{A}_2)_{\hat{\rho}_b}+\mathrm{Var}(\hat{A}_3)_{\hat{\rho}_c}], \nonumber \\
           B_2=4 \,\mathrm{max}\{&\mathrm{Var}(\hat{A}_1)_{\hat{\rho}_a}+\mathrm{Var}(\hat{A}_2+\hat{A}_3)_{\hat{\rho}_{bc}},\nonumber\\
    &\mathrm{Var}(\hat{A}_2)_{\hat{\rho}_b}+\mathrm{Var}(\hat{A}_1+\hat{A}_3)_{\hat{\rho}_{ac}},\nonumber\\
   &\mathrm{Var}(\hat{A}_3)_{\hat{\rho}_c}+\mathrm{Var}(\hat{A}_1+\hat{A}_2)_{\hat{\rho}_{ab}}
   \}, \nonumber\\
       B_3=4\, [\mathrm{Var}(&\hat{A}_1+\hat{A}_2+\hat{A}_3)_{\hat{\rho}}]. \nonumber
\end{align}
If the left-hand side $F_Q[\hat{\rho}_{\mathrm{sep}},\hat{A}]>B_1$, then the state is not separable, i.e. there is entanglement between three modes. Furthermore, if $F_Q[\hat{\rho}_{\mathrm{sep}},\hat{A}]>B_2$, the state is fully inseparable, which means full tripartite inseparability between three modes. Furthermore, $B_3$ give a bound valid for all physical states, which satisfy $F_Q[\hat{\rho},\hat{A}]\leq B_3$ for arbitrary density matrices $\hat{\rho}$.

The metrological witness for entanglement depends on the choice of the local operator $\hat{A}$. Certain choices of operators may be better suited than others to detect entanglement in a given state $\hat{\rho}$. In order to find an optimal operator for a family of accessible $\mathbf{\hat{A}}_j=[\hat{A}_j^{(1)},\hat{A}_j^{(2)}\cdots]^T$, the local operators $\hat{A}_j$ are denoted by the expression $\sum_{m=1}c_j^{(m)}\hat{A}_j^{(m)}=\mathbf{c}_j\cdot\mathbf{\hat{A}}_j$, and the full generator of the unitary transformation $\hat{A}(\mathbf{c})=\mathbf{c}_1\cdot \mathbf{\hat{A}}_1+\mathbf{c}_2\cdot \mathbf{\hat{A}}_2+\mathbf{c}_3\cdot \mathbf{\hat{A}}_3$ is characterized by the vector $\mathbf{c}=[\mathbf{c}_1,\mathbf{c}_2,\mathbf{c}_3]^T$. According to Eq.~\ref{eqfv1}, the quantity
\begin{equation}
    W[\hat{\rho},\hat{A}(\mathbf{c})]=F_Q[\hat{\rho},\hat{A}(\mathbf{c})]-4\sum_{j=1}^N \mathrm{Var}(\mathbf{c}_j\cdot\mathbf{\hat{A}}_j)_{\hat{\rho}}
\end{equation}
must be nonpositive for arbitrary choices of $\mathbf{c}$ whenever the state $\rho$ is separable. We can now maximize $W[\hat{\rho},\hat{A}(\mathbf{c})]$ by variation of $\mathbf{c}$ to obtain an optimized entanglement witness for the state $\rho$, given the sets of available operators contained in $\mathcal{A}=\{\mathbf{\hat{A}}_1,\mathbf{\hat{A}}_2,\mathbf{\hat{A}}_3\}$.

To this aim, let us first express the quantum Fisher information in matrix form as $F_Q[\hat{\rho},\hat{A}(\mathbf{c})]=\mathbf{c}^T Q_{\rho}^{\mathcal{A}}c$, where the spectral decomposition $\hat{\rho}=\sum_k p_k|\Psi_k\rangle\langle\Psi_k|$ defines $(Q_{\hat{\rho}}^{\mathcal{A}})_{i j}^{m n}=$ $2 \sum_{k, l} \frac{(p_{k}-p_{l})^{2}}{p_{k}+p_{l}}\langle\Psi_{k}|\hat{A}_{i}^{(m)}| \Psi_{l}\rangle\langle\Psi_{l}|\hat{A}_{j}^{(n)}| \Psi_{k}\rangle$ element-wise and the sum extends over all pairs with $p_k+p_l\neq 0$. The indices $i$ and $j$ refer to different parties, while the indices $m$ and $n$ label the respective local sets of observables. Similarly, we can express the  elements of the covariance matrix of $\hat{\rho}$ as $(\Gamma_{\hat{\rho}}^{\mathcal{A}})_{i j}^{m n}=\mathrm{Cov}(\hat{A}_{i}^{(m)}, \hat{A}_{j}^{(m)})_{\hat{\rho}}$. If the above covariance matrix is evaluated after replacing $\hat{\rho}$ with $\Pi(\hat{\rho})=\hat{\rho}_{1} \otimes \cdots \otimes \hat{\rho}_{N}$, where $\hat{\rho}_i$ is the reduced density operator, we arrive at the expression for the local variances, $\sum_{j=1}^{N} \mathrm{Var}(\mathbf{c}_{j}\cdot \hat{\mathbf{A}}_{j})_{\hat{\rho}}=\mathbf{c}^{T} \Gamma_{\Pi(\hat{\rho})}^{\mathcal{A}} \mathbf{c}$. Combining this with expression for the quantum Fisher matrix, the separability criterion reads 
\begin{equation}\label{eqW1}
W[\hat{\rho}, \hat{A}(\mathbf{c})]=\mathbf{c}^{T}\left(Q_{\hat{\rho}}^{\mathcal{A}}-4 \Gamma_{\Pi(\hat{\rho})}^{\mathcal{A}}\right) \mathbf{c} \leq0.
\end{equation}
An entanglement witness is therefore found when the matrix $\mathbf{c}^{T}(Q_{\hat{\rho}}^{\mathcal{A}}-4 \Gamma_{\Pi(\hat{\rho})}^{\mathcal{A}}) \mathbf{c}$ has at least one positive eigenvalue. The criterion (\ref{eqW1}) can be equivalently stated as $\lambda_{\mathrm{max}}(Q_{\hat{\rho}}^{\mathcal{A}}-4 \Gamma_{\Pi(\hat{\rho})}^{\mathcal{A}}) \leq 0$, where $\lambda_{max}(M)$ denotes the largest eigenvalue of the matrix $M$.

For pure states $\hat{\rho}=|\Psi\rangle\langle\Psi|$, the quantum Fisher matrix coincides, up to a factor of 4 , with the covariance matrix, i.e., $Q_{|\Psi\rangle}^{\mathcal{A}}=4 \Gamma_{|\Psi\rangle}^{\mathcal{A}}$. Thus, according to Eq.~(\ref{eqW1}), every pure separable state must satisfy the condition
\begin{equation}
 \Gamma_{|\Psi\rangle}^{\mathcal{A}}-\Gamma_{\Pi(\mid \Psi))}^{\mathcal{A}} \leq 0.  
\end{equation}

A common choice for such a set are the local position operators $x_j$ and momentum operators $p_j$. As our state is non-Gaussian, their characteristics cannot be sufficiently uncovered by measuring linear observables. We need extend the family of accessible operators by adding second order nonlinear local operators: $\mathbf{\hat{A}}_j=[\hat{x}_j,\hat{p}_j,\hat{x}_j^2,\hat{p}_j^2,(\hat{x}_j\hat{p}_j+\hat{p}_j\hat{x}_j)/2]^T$.
The optimum is given by the maximum eigenvalue of the matrix $(Q_{\hat{\rho}}^{\mathcal{A}}-4 \Gamma_{\Pi(\hat{\rho})}^{\mathcal{A}}) $, the optimal direction is given by the corresponding eigenvector. The result is $\mathbf{c}_j=[0,0,1,1,0]$, Which indicates the optimal operators taking the form: $\hat{A}_j^{opt}=\hat{x}_j^2+\hat{p}_j^2$. And the full optimal operator is $\hat{A}_{opt}=\sum_{j=1}^3 \hat{A}_j^{opt}=\hat{x}_1^2+\hat{p}_1^2+\hat{x}_2^2+\hat{p}_2^2+\hat{x}_3^3+\hat{p}_3^2=4(\hat{a}_1^{\dagger}\hat{a}_1+\hat{a}_2^{\dagger}\hat{a}_2+\hat{a}_3^{\dagger}\hat{a}_3)+6$, where we take
$\hat{x}_i=\hat{a}_j+\hat{a}_j^{\dagger}$,  $\hat{p}_j=-i(\hat{a}_j-\hat{a}_j^{\dagger})$ for $j=1,2,3$.
As the constant in the operator $\hat{A}$ does not influence the optimal result, so the optimal operator can be rewritten in the form of $\hat{A}_{opt}=\hat{N}=\hat{N}_1+\hat{N}_2+\hat{N}_3$. 

\section{The optimal squeezing parameter}\label{optsqueezing}
\subsection{Squeezing parameter}
The Fisher information has a lower bound, which is given by
\begin{equation}
 \chi^{-2}[\hat{\rho}(\theta),\hat{M}] \leq F[\hat{\rho}(\theta),\hat{M}] \leq F_{Q}[\hat{\rho}, \hat{A}],   
\end{equation}
where $\chi^{-2}[\hat{\rho}(\theta),\hat{M}]$ is the reciprocal of the squeezing parameter 
and $\hat{\rho}({\theta})=e^{-i \hat{A} \theta} \hat{\rho} e^{i \hat{A} \theta}$ is obtained by applying an unitary evolution $\hat{A}$ on initial state $\hat{\rho}$.
And the chain of inequalities is saturable by an optimal measurement observable $\hat{M}$. The squeezing parameter is also used to estimate an unknown phase $\theta$ encoded in a quantum state $\hat{\rho(\theta)}$ by the method of moments (see Ref.~\cite{RN01} for details). In the central limit ($n\gg1)$, the phase uncertainty is given  by $\left(\Delta \theta_{\text {est }}\right)^{2}=\chi^{2}[\hat{\rho}(\theta), \hat{M}] / n$, where $\chi^{2}[\hat{\rho}(\theta), \hat{M}]=$ $(\Delta \hat{M})_{\hat{\rho}(\theta)}^{2}\left(d\langle\hat{M}\rangle_{\hat{\rho}(\theta)} / d \theta\right)^{-2}$ is the squeezing parameter of $\hat{\rho}$ associated with the measurement of the observable $\hat{M}$. Specially, for unitary evolution $\hat{\rho}({\theta})=e^{-i \hat{A} \theta} \hat{\rho} e^{i \hat{A} \theta}$, the squeezing parameter is a property of the initial state $\hat{\rho}$, which takes the form of,
\begin{equation}
   \chi^2[\hat{\rho},\hat{A},\hat{M}]=\frac{(\Delta\hat{M})_{\hat{\rho}}^2}{\left|\langle[\hat{A}, \hat{M}]\rangle_{\hat{\rho}}\right|^{2}},
\end{equation}
with the unitary evolution $\hat{A}$ and observable $\hat{M}$.

\subsection{The Optimal observable in nonlinear quadrature parameters}

\begin{figure}[t]
	\centering
	\includegraphics[scale=0.42]{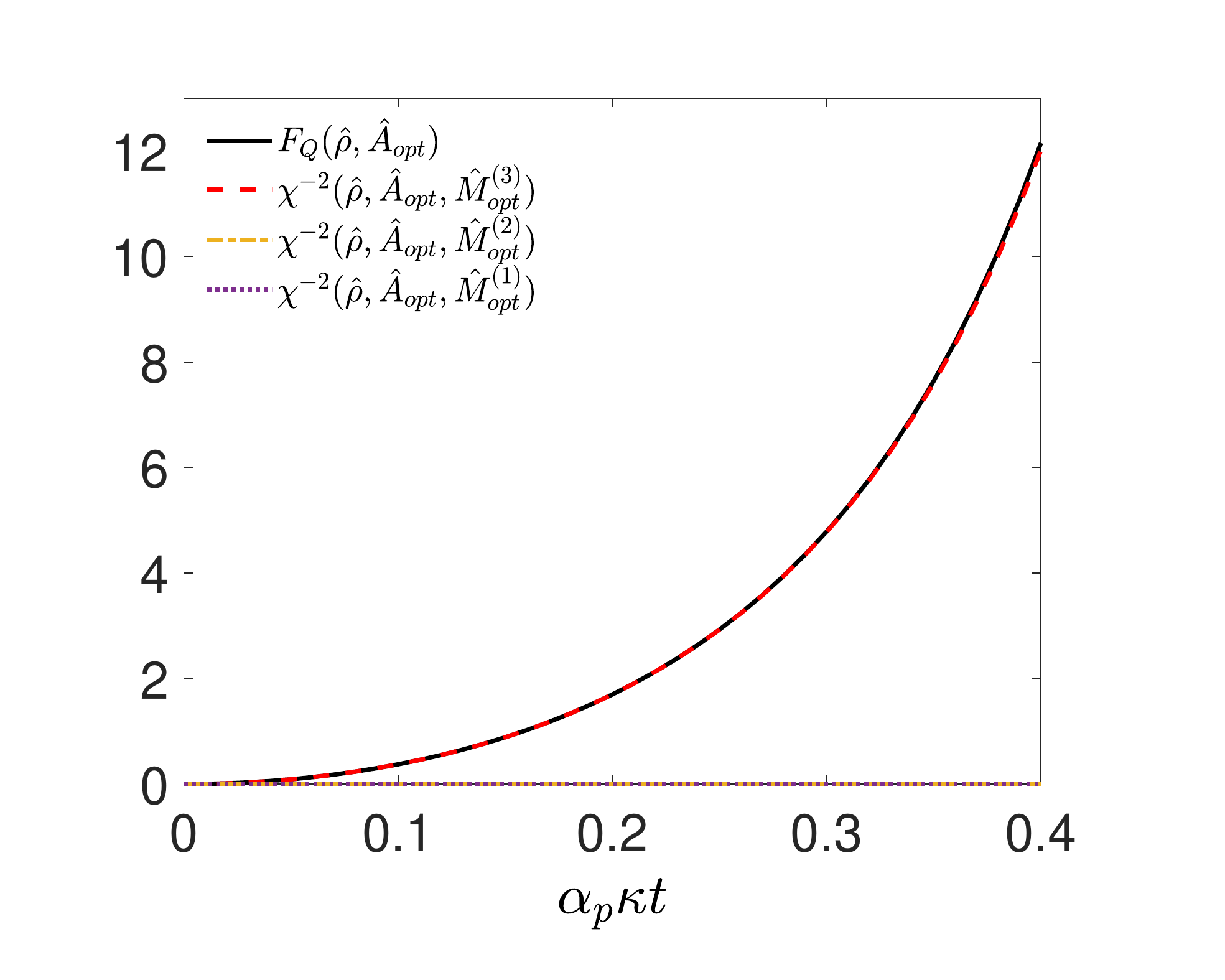}
	\caption{Evolution of the optimal nonlinear parameters $\chi^{-2}$ are plotted as a function of the effective coupling strength $\alpha_p\kappa t$ by taking different order correlations in $\hat{M}$.
	Here, the solid red, yellow, and dashed purple lines denote the first-, second-, and third-order correlations, respectively.
    The black solid line denotes the QFI.
	}
	\label{sfig1}
\end{figure}

The analytical optimization is over arbitrary linear combinations of accessible operators $(\hat{M}=\mathbf{n}\cdot \mathbf{\hat{M}}=\sum_l n_l \hat{M}_l)$, where $\mathbf{\hat{M}}$ include the different observable.
If we only consider the first-order operators, the set $\mathbf{M}$ can be written:
\begin{equation}
    \mathbf{\hat{M}^{(1)}}=[\hat{x}_1,\hat{p}_1,\hat{x}_2,\hat{p}_2,\hat{x}_3,\hat{p}_3],
\end{equation}
And if we add the second order operators, the set $\mathbf{\hat{M}}$ can be expanded over the following 27 terms
\begin{equation}
\begin{aligned}
    \mathbf{\hat{M}^{(2)}}=\{
    &\mathbf{\hat{M}^{(1)}},[\hat{x}_1^2,\hat{p}_1^2,\hat{x}_2^2,\hat{p}_2^2,\hat{x}_3^2,\hat{p}_3^2,\hat{x}_1\hat{p}_1,\hat{x}_1 \hat{x}_2,\\
    &\hat{x}_1 \hat{p}_2,\hat{x}_1\hat{x}_3,\hat{x}_1\hat{p}_3,
   \hat{x}_2\hat{p}_1,\hat{p}_1\hat{p}_2,\hat{x}_3\hat{p}_1,\hat{p}_1\hat{p}_3,\\
   &\hat{x}_2\hat{p}_2,\hat{x}_2\hat{x}_3,\hat{x}_2\hat{p}_3,\hat{p}_2\hat{x}_3,\hat{p}_2\hat{p}_3,\hat{x}_3\hat{p}_3]\} \;,      
\end{aligned}
\end{equation}
if we add the third order operators, the set can be written as the 83 terms
\begin{equation}
\begin{aligned}
\hat{\mathbf{M}}^{(3)}=&\{\hat{\mathbf{M}}^{(1)}, \hat{\mathbf{M}}^{(2)},[\hat{x}_{1}^{3}, \hat{p}_{1}^{3}, \hat{x}_{2}^{3}, \hat{p}_{2}^{3}, \hat{x}_{3}^{3}, \hat{p}_{3}^{3}, \hat{x}_{1}^{2} \hat{p}_{1},\\
& \hat{x}_{1}^{2} \hat{x}_{2}, \hat{x}_{1}^{2} \hat{p}_{2}, \hat{x}_{1}^{2} \hat{x}_{3}, \hat{x}_{1}^{2} \hat{p}_{3}, \hat{p}_{1}^{2} \hat{x}_{1}, \hat{p}_{1}^{2} \hat{x}_{2}, \hat{p}_{1}^{2} \hat{p}_{2}, \\
& \hat{p}_{1}^{2} \hat{x}_{3}, \hat{p}_{1}^{2} \hat{p}_{3}, \hat{x}_{2}^{2} \hat{x}_{1}, \hat{x}_{2}^{2} \hat{p}_{1}, \hat{x}_{2}^{2} \hat{p}_{2}, \hat{x}_{2}^{2} \hat{x}_{3}, \hat{x}_{2}^{2} \hat{p}_{3}, \\
& \hat{p}_{2}^{2} \hat{x}_{1}, \hat{p}_{2}^{2} \hat{p}_{1}, \hat{p}_{2}^{2} \hat{x}_{2}, \hat{p}_{2}^{2} \hat{x}_{3}, \hat{p}_{2}^{2} \hat{p}_{3}, \hat{x}_{3}^{2} \hat{x}_{1}, \hat{x}_{3}^{2} \hat{p}_{1}, \\
& \hat{x}_{3}^{2} \hat{x}_{2}, \hat{x}_{3}^{2} \hat{p}_{2}, \hat{x}_{3}^{2} \hat{p}_{3}, \hat{p}_{3}^{2} \hat{x}_{1}, \hat{p}_{3}^{2} \hat{p}_{1}, \hat{p}_{3}^{2} \hat{x}_{2}, \hat{p}_{3}^{2} \hat{p}_{2}, \\
& \hat{p}_{3}^{2} \hat{x}_{3}, \hat{x}_{1} \hat{x}_{2} \hat{x}_{3}, \hat{x}_{1} \hat{x}_{2} \hat{p}_{1}, \hat{x}_{1} \hat{x}_{2} \hat{p}_{2}, \hat{x}_{1} \hat{x}_{2} \hat{p}_{3}, \\
& \hat{x}_{1} \hat{x}_{3} \hat{p}_{1}, \hat{x}_{1} \hat{x}_{3} \hat{p}_{2}, \hat{x}_{1} \hat{x}_{3} \hat{p}_{3}, \hat{x}_{1} \hat{p}_{1} \hat{p}_{2}, \hat{x}_{1} \hat{p}_{1} \hat{p}_{3}, \\
& \hat{x}_{1} \hat{p}_{2} \hat{p}_{3}, \hat{x}_{2} \hat{x}_{3} \hat{p}_{1}, \hat{x}_{2} \hat{x}_{3} \hat{p}_{2}, \hat{x}_{2} \hat{x}_{3} \hat{p}_{3}, \hat{x}_{2} \hat{p}_{1} \hat{p}_{2}, \\
& \hat{x}_{2} \hat{p}_{1} \hat{p}_{3}, \hat{x}_{2} \hat{p}_{2} \hat{p}_{3}, \hat{x}_{3} \hat{p}_{1} \hat{p}_{2}, \hat{x}_{3} \hat{p}_{1} \hat{p}_{3}, \hat{x}_{3} \hat{p}_{2} \hat{p}_{3}, \\
&\hat{p}_{1} \hat{p}_{2} \hat{p}_{3}]\} \;.
\end{aligned}
\end{equation}
From these we obtain:
\begin{equation}
\begin{aligned}
      \chi^{-2}[\hat{\rho},\hat{A}_{opt},\hat{M}]&=\frac{\left|\langle[\hat{A}_{opt}, \hat{M}]\rangle_{\hat{\rho}}\right|^{2}}{\operatorname{Var}[ \hat{M}]_{\hat{\rho}}}\\
      &=\mathbf{n}^T\mathbf{C}[\hat{\rho},\mathbf{\hat{M}}]\mathbf{\Gamma}[\hat{\rho},\mathbf{\hat{M}}]^{-1}\mathbf{n},    
\end{aligned}
\end{equation}
where $\mathbf{\Gamma}[\hat{\rho},\mathbf{\hat{M}}]_{kl}=\operatorname{Cov}(\hat{M}_k,\hat{M}_l)$ and $\mathbf{C}[\hat{\rho},\mathbf{\hat{M}}]_{kl}=\langle[\hat{A}_{opt},\hat{M}_k]\rangle \langle[\hat{M}_l,\hat{A}_{opt}]\rangle$. The optimum is given by the maximum eigenvalue of the matrix $\mathbf{C}[\hat{\rho},\mathbf{\hat{M}}]\mathbf{\Gamma}[\hat{\rho},\mathbf{\hat{M}}]^{-1} $, and the optimal direction is given by the corresponding eigenvector. We show the results in Fig.~\ref{sfig1}, where we find that the optimal value is zero with only considering first and second order observables. Instead, if we add third-order measurements, the optimum of the nonlinear parameter $\chi^{-2}[\hat{\rho},\hat{A}_{opt},\hat{M}_{opt}]$ can be nearly equal to the QFI $F_Q[\hat{\rho},\hat{A}_{opt}]$, which indicates the $\hat{M}$ in third-order is the optimal measurement in this case. The optimal measurement takes the form
\begin{equation}
  \hat{M}_{opt}=\hat{x}_1 \hat{x}_2 \hat{x}_3-\hat{x}_1 \hat{p}_2 \hat{p}_3-\hat{p}_1 \hat{x}_2 \hat{p}_3-\hat{p}_1 \hat{p}_2 \hat{x}_3.
\end{equation}

\section{Bound of metrological useful resource}\label{C}

It has been clarified that nonclassicality is a necessary resource to achieve quantum advantage in quantum metrology tasks~\cite{PhysRevLett.122.040503}. Therefore, we will try to distinguish the useful resource for quantum metrology with the help of the nonclassicality. 

To this aim, we will take use of the following two properties of QFI:
(i) For pure states, such as coherent states $|\alpha\rangle$, the quantum Fisher information becomes proportional to the variance of the generator:
\begin{equation}
\begin{aligned}
  F_{Q}(|\alpha\rangle, \hat{G})&=4\left(\Delta_{\alpha} \hat{G}\right)^{2}\\
  &=4\left(\left\langle\alpha\left|\hat{G}^{2}\right| \alpha\right\rangle-\langle\alpha|\hat{G}| \alpha\rangle^{2}\right).       
\end{aligned}
\end{equation}
(ii) The quantum Fisher information is convex. Thus, for classical states
\begin{equation}
\rho_{\text {class }}=\int d^{2} \alpha P_{\text {class }}(\alpha)|\alpha\rangle\langle\alpha|,
\end{equation}
where $P_{\text {class }}(\alpha)$ is a non-negative function no more singular than a delta function.

Based on these properties, the bound for the quantum Fisher information of classical states reads~\cite{RN612}:
\begin{equation}\label{nonclass}
F_{Q}\left(\rho_{\text {class }}, \hat{G}\right)  \leq \int d^{2} \alpha P_{\text {class }}(\alpha) F_{Q}(|\alpha\rangle, \hat{G}).    
\end{equation}
Taking the generator $\hat{G}=\hat{a}^{\dagger}\hat{a}$ into account, the right-hand side of inequality~(\ref{nonclass}) is then given by its means number of photons in coherent states:
\begin{equation}
\begin{aligned}
      F_{Q}\left(\left|\alpha_{\hat{\rho}}\right\rangle, \hat{a}^{\dagger} \hat{a}\right)&=4\Delta_{\alpha}^{2} (\hat{a}^{\dagger} \hat{a})\\
    &=4\left\langle\alpha_{\hat{\rho}}\left|\hat{a}^{\dagger} \hat{a}\right| \alpha_{\hat{\rho}}\right\rangle\\
    &=4 \operatorname{Tr}\left(\hat{\rho} \hat{a}^{\dagger} \hat{a}\right).
\end{aligned}
\end{equation}
So Eq.~(\ref{nonclass}) can be rewritten for classical states as
\begin{equation}
     F_Q(\hat{\rho},\hat{a}^{\dagger}\hat{a})\leq 4Tr(\hat{\rho}\hat{a}^{\dagger}\hat{a}). 
 \end{equation}
This approach is also suitable for a classical multimode system with the generators $\hat{G}=\hat{N}=\hat{a}_1^{\dagger}\hat{a}_1+\hat{a}_2^{\dagger}\hat{a}_2+\hat{a}_3^{\dagger}\hat{a}_3$, as the QFI for separable coherent states is given by:
\begin{equation}
\begin{aligned}
F_{Q}\left(\left|\alpha_1\alpha_2\alpha_3\right\rangle_{\hat{\rho}}, \hat{N}\right)&=
4\left(\Delta_{\hat{\rho}} \hat{N}\right)^{2}\\
&=
4\left\langle\alpha_1\alpha_2\alpha_3\left|\hat{N}\right| \alpha_1\alpha_2\alpha_3\right\rangle_{\hat{\rho}}\\
&=4 \operatorname{tr}\left(\hat{\rho} \hat{N}\right).      
\end{aligned}
\end{equation}
So the bound for QFI of classical multimode system reads:
\begin{equation}\label{fqn}
     F_Q(\hat{\rho},\hat{N})\leq 4Tr(\hat{\rho}\hat{N}). 
 \end{equation}
The violation of Eq.~(\ref{fqn}) means the state is nonclassical, which indicates the state is a useful resource for quantum metrology.

\section{The simplified Squeezing Parameter}\label{simplesqueezing}
In the above, we obtained the optimal nonlinear squeezing parameter $\chi^{-2}(\hat{\rho},\hat{A}_{opt},\hat{M}_{opt})$, which requires at least four observables for $\hat{M}_{opt}$ and four for $[\hat{A}_{opt},\hat{M}_{opt}]$ in the three-mode SPDC systems:
\begin{align}
      \hat{M}_{opt}=&\hat{x}_1\hat{x}_2\hat{x}_3-\hat{x}_1\hat{p}_2\hat{p}_3-\hat{p}_1 \hat{x}_2 \hat{p}_3-\hat{p}_1\hat{p}_2\hat{x}_3,\nonumber\\
    [\hat{A}_{opt},\hat{M}_{opt}]=&3i(\hat{p}_1\hat{x}_2\hat{x}_3+\hat{x}_1\hat{p}_2\hat{x}_3+\hat{x}_1\hat{x}_2\hat{p}_3\\
    &-\hat{p}_1\hat{p}_2\hat{p}_3). \nonumber
\end{align}
In the following, we want to simplify the squeezing parameter to be more easily accessible in experiments.
One collective measurement can be written as
$\hat{Q}_j=[\hat{x}_1\sin(\theta_1^j)+\hat{p}_1\cos(\theta_1^j)]\times[\hat{x}_2\sin(\theta_2^j)+\hat{p}_2\cos(\theta_2^j)]\times[\hat{x}_3\sin(\theta_3^j)+\hat{p}_3\cos(\theta_3^j)]$, where $\theta_j\in [0,2\pi)$. 
Our aim is to use fewer collective measurements to obtain the squeezing parameter that can be used to witness entanglement.
To obtain the optimal simplified squeezing parameter, we need to consider both $\hat{M}$ and $[\hat{A}_{opt},\hat{M}]$, where $\hat{M}$ can be written in the form of $\hat{M}=\sum_i^n \hat{Q}_j$ with n measurements for $\hat{M}$.
By optimizing the parameter $\theta_i$ in $\hat{M}$  with  fixed number of collective measurements, we can find that  
we need at least 1 observable in $\hat{M}$. There are four optimal observables of $\hat{M}$, which are shown in the following:
\begin{equation}
  \begin{aligned}
    \hat{M}_1^{(1)}&=\hat{x}_1\hat{x}_2\hat{x}_3,\\
    [\hat{A}_{opt},\hat{M}_1^{(1)}]&=i(\hat{p}_1\hat{x}_2\hat{x}_3+\hat{x}_1\hat{p}_2\hat{x}_3+\hat{x}_1\hat{x}_2\hat{p}_3);\\
    \hat{M}_1^{(2)}&=\hat{x}_1\hat{p}_2\hat{p}_3,\\
    [\hat{A}_{opt},\hat{M}_1^{(2)}]&=i(\hat{p}_1\hat{p}_2\hat{p}_3-\hat{x}_1\hat{x}_2\hat{p}_3-\hat{x}_1\hat{p}_2\hat{x}_3); \\
    \hat{M}_1^{(3)}&=\hat{p}_1\hat{x}_2\hat{p}_3,\\
    [\hat{A}_{opt},\hat{M}_1^{(3)}]&=i(\hat{p}_1\hat{p}_2\hat{p}_3-\hat{x}_1\hat{x}_2\hat{p}_3-\hat{p}_1\hat{x}_2\hat{x}_3); \\
    \hat{M}_1^{(4)}&=\hat{p}_1\hat{p}_2\hat{x}_3,\\
    [\hat{A}_{opt},\hat{M}_1^{(4)}]&=i(\hat{p}_1\hat{p}_2\hat{p}_3-\hat{p}_1\hat{x}_2\hat{x}_3-\hat{x}_1\hat{p}_2\hat{x}_3).
\end{aligned}  
\end{equation}

In the main text, we have set $\hat{M}_1=\hat{M}_1^{(1)}$ and obtained the corresponding results. Please note that all of the simplified nonlinear parameters $\chi^{-2}(\hat{M}_1)$ behave the same when choosing $\hat{M}_1$ as each of the four optimal observables. 

With considering two observables in $\hat{M}$, the optimal $\hat{M}$ is given by any two terms out of four terms in $\hat{M}_{opt}$, for example, we show one of the optimal results is: 
\begin{align}
      \hat{M}_2=&\hat{x}_1\hat{x}_2\hat{x}_3-\hat{x}_1\hat{p}_2\hat{p}_3,\nonumber\\
    [\hat{A}_{opt},\hat{M}_2]=&i(\hat{p}_1\hat{x}_2\hat{x}_3+2\hat{x}_1\hat{p}_2\hat{x}_3+2\hat{x}_1\hat{x}_2\hat{p}_3\\
    &-\hat{p}_1\hat{p}_2\hat{p}_3). \nonumber
\end{align}

When considering three observables, the optimal $M$ is given by any three terms out of four terms in $M_{opt}$, for example, we show one of the optimal results is: 
\begin{align}
      \hat{M}_3=&\hat{x}_1\hat{x}_2\hat{x}_3-\hat{x}_1\hat{p}_2\hat{p}_3-\hat{p}_1 \hat{x}_2 \hat{p}_3,\nonumber\\
    [\hat{A}_{opt},\hat{M}_3]=&i(2\hat{p}_1\hat{x}_2\hat{x}_3+2\hat{x}_1\hat{p}_2\hat{x}_3+3\hat{x}_1\hat{x}_2\hat{p}_3\\&-2\hat{p}_1\hat{p}_2\hat{p}_3).\nonumber
\end{align}
The results of the simplified nonlinear parameters with different observables are shown in Fig.~\ref{fig1}.


%

\end{document}